\documentclass[12pt]{article}

\usepackage{amsfonts,amssymb,amsmath}

\usepackage{graphicx}
\DeclareGraphicsExtensions{.pdf,.png,.jpg}

\begin{document}
	
\title{\bf Quantum engine based on general measurements}
\author{ Naghi Behzadi $^{a}$ \thanks{E-mail:n.behzadi@tabrizu.ac.ir} ,
	\\ $^a${\small Department of theoretical physics and astrophysics, University of Tabriz, 5166614776, Tabriz, Iran.}\\ {\small }
} \maketitle

\begin{abstract}
In this work, we introduce a three-stroke quantum engine with a single-qubit working substance whose cycle consists of two strokes arise due to performing two distinct general quantum measurements and it is completed by thermalization through contact with a finite temperature thermal reservoir. It is demonstrated that energy is imported into the engine by first measurement channel and work (useful energy) is extracted from it, in a frictionless isentropic manner, by the second measurement channel. It is particularly shown that the engine is capable to have perfect efficiency. In continuation, we equip the the engine cycle with two additional adiabatic strokes. As illustrated, the presence of adiabatic strokes in the cycle provide an improved lower bound for the parametrized efficiency range of the engine.
\end{abstract}

\section{Introduction}

The task of a heat engine is conversion some form of stochastic energy into mechanical work or organized energy, in a cyclically repeated process. In a traditional heat engine, quantum or classical, the mechanical work is extracted through a process that the engine is enabled to gain an amount of heat (unorganized energy) from a hot thermal reservoir and loses a less value of it into a cold reservoir [1-3]. This arises from second law of thermodynamics that states it is impossible conversion of heat to work perfectly \cite{planck}. In fact, in order to restore the initial state of working substance and complete the cycle, dissipation of finite amount of thermal energy into the cold reservoir is unavoided. 

However, there is another type of quantum engine operating at a finite temperature thermal reservoir and assisted by a so-called Maxwell's demon [4-13]. The demon exploits information, obtained from working substance through projective measurements, to extract work from finite temperature thermal reservoir. This process is known as work extraction through feed back control [14-16]. The appearance contradiction to the second law can be resolved if the energy cost of reseting the demon memory is included.

As is known, projective measurements on a quantum system are able to increase its average energy and entropy, provided that the measured observable does not commute with the Hamiltonian of that system [17-20]. Therefore, projective measurement can be regarded by itself as a fuel in a quantum engine. In other words, projective measurements play the corresponding role of a hot thermal reservoir in traditional quantum heat engines \cite{talkner}. In Ref \cite{elouard1}, for a typical Maxwell's demon engine, the thermal bath, as a stochastic energy source, is replaced by projective measurements performed by the demon. Also in this direction, Ref \cite{elouard2} proposes measurement fueled quantum engines based on position-resolving measurement performed on a quantum particle and work is extracted by moving the particle against a potential barrier.

In this paper, a quantum engine is introduced whose cyclically working process consists of two strokes established by exploiting two set of distinct general quantum measurements \cite{jacobs,wiseman}, along with thermalization through a finite temperature thermal reservoir. We stress that, for this engine, the work extraction process does not essentially involve a time dependent Hamiltonian. Two strokes of proposed general measurements can be seen as thermodynamics resources analogous to heat and work reservoirs, and thermalization process through contact with a thermal reservoir completes the cycle. It is emphasized that the second stroke takes place in an isentropic fashion to make the work extraction as efficient as possible. Here, the isentropic process takes place without quantum friction, i.e., without population of nondiagonal elements of density matrix (quantum coherence). We use controllable property of proposed general measurements to supply the engine such that its efficiency can be approached to unit. In addition, we extend our engine by including two additional adiabatic strokes into the cycle. As illustrated in the text, we parametrize the engine efficiency in terms of measurement strength parameter of the first stroke. In this step, it is shown that the efficiency range of the engine, the range in which the engine efficiency varies in terms of the introduced parameter, experiences an improved lower bound in comparison to the previous case. Interestingly, for a special value of the parameter, our engine gives the same efficiency as of the engine introduced in \cite{talkner}.

The paper is organized as follows: In Sec. II, the structure and performance of the quantum engine on the basis of proposed general quantum measurements are illustrated. Sec. III is devoted to extend the introduced engine in Sec. II, by including two additional adiabatic strokes into the engine cycle. Finally, the paper is ended by a brief conclusion in Sec. IV.

\section{Modeling the quantum engine}

We consider a quantum engine consisting of a single-qubit working substance undergoing a three-stroke cycle. The qubit is initially at thermal equilibrium with a thermal reservoir at temperature $T$. It is assumed that the qubit has following Hamiltonian
\begin{eqnarray}
H=\frac{\hbar\omega_{0}}{2}{\sigma_{z}},
\end{eqnarray}   
where $\omega_{0}$ is transition frequency of the qubit, and $\sigma_{z}$ is the $z$-component Pauli spin operator. The corresponding  thermalized (Gibbs) state of the qubit is  
\begin{eqnarray}
\rho^{th}=\frac{e^{-\beta H}}{Z(\beta)},
\end{eqnarray}
where $\beta$ is the relative inverse temperature, and $Z(\beta)=\mathrm{Tr}(e^{-\beta H})=2\mathrm{cosh}(\beta\hbar\omega_{0}/2)$ is the canonical partition function. Therefore, the average thermal energy of the system is
\begin{eqnarray}
\mathcal{E}^{th}=\mathrm{Tr}(H\rho^{th})=-\frac{\hbar\omega_{0}}{2}\mathrm{tanh}\left(\frac{\beta\hbar\omega_{0}}{2}\right).
\end{eqnarray}

In the next step, it is assumed that the engine is isolated from the thermal reservoir and the working substance of the engine undergoes strokes due to performing general measurements. Generally speaking \cite{jacobs}, to perform general measurement on a quantum system whose pre-measurement state is $\rho$, it should be allowed to interact through unitary dynamics with another quantum system called probe or measurement device. After interaction, density operator of the combined system is $\rho^{U}=U\left(\rho\otimes|0\rangle\langle0|\right)U^{\dagger}$, where $\rho\otimes|0\rangle\langle0|$ is initial state of the system-probe and $U$ is any unitary operator acting on the combined system. It is usually assumed, without lose of generality, that the initial state of the probe is $|0\rangle\langle0|$, and it is not correlated with the system initially. The subsequent von Neumann measurement on the probe, projects it onto one of its eigenstates, namely $|n\rangle$, with following probability 
\begin{eqnarray}
p_{n}=\mathrm{Tr}\left((|n\rangle\langle n|\otimes I)\rho^{U}(|n\rangle\langle n|\otimes I)\right)
=\mathrm{Tr}\left(|n\rangle\langle n|\otimes A_{n}\rho A_{n}^{\dagger}\right)=\mathrm{Tr}\left(A_{n}^{\dagger}A_{n}\rho \right),
\end{eqnarray} 
where $A_{n}=\langle n|U|0\rangle$, so the corresponding normalized post-measurement state for the system becomes 
\begin{eqnarray}
\tilde{\rho}=\frac{A_{n}\rho A_{n}^{\dagger}}{\mathrm{Tr}\left(A_{n}\rho A_{n}^{\dagger} \right)}.
\end{eqnarray} 
Therefore, we have the unselective post-measurement state for the system as 
\begin{eqnarray}
\rho'=\sum_{n}^{N}A_{n}\rho A_{n}^{\dagger},
\end{eqnarray} 
where $N$ is the number of possible measurement outcomes.
Since the sum of probability of outcomes must be unit in equation (4), then it is concluded that
\begin{eqnarray}
\sum_{n}^{N}A_{n}^{\dagger}A_{n}=I.
\end{eqnarray} 
In fact, equation (7) is the essence of fundamental theorem of quantum measurement \cite{jacobs}, which states that every set of operators $\{A_{n}\}$, $n=1, 2, ..., N$, that satisfies equation (7), describes a measurement on the quantum system with outcomes as $\tilde{\rho}$ in (5), and corresponding probabilities as $p_{n}$ in (4).  

Now we see that how the quantum engine, after thermal isolation, undergoes thermodynamics strokes caused by measurement channels. In order to elucidate the first stroke, consider following operators for the single-qubit engine as 
\begin{eqnarray}
M_{1}(\mathcal{P})=\sqrt{1-\mathcal{P}}|0\rangle\langle 0|+|1\rangle\langle 1|, \quad M_{2}(\mathcal{P})=\sqrt{\mathcal{P}}|1\rangle\langle 0|,
\end{eqnarray} 
where $\mathcal{P}$ is the measurement strength parameter. Operator $M_{1}(\mathcal{P})$ is similar to the weak quantum measurement reversal map of a qubit with strength $\mathcal{P}$ in the computational basis $\{|0\rangle,|1\rangle\}$ \cite{kim,he,behzadi}. Obviously, these operators satisfies equation (7), so they describe a quantum measurement on the working substance. In addition, the operators
\begin{eqnarray}
E_{1}(\mathcal{P}):=M_{1}^{\dagger}M_{1}=(1-\mathcal{P})|0\rangle\langle 0|+|1\rangle\langle 1|, \quad E_{2}(\mathcal{P}):=M_{2}^{\dagger}M_{2}=\mathcal{P}|0\rangle\langle 0|,
\end{eqnarray}  
are the so-called $POVM$ (positive operator valued measure) operators corresponding to the measurement operators (8). It is clear that, $0 \leq\mathcal{P} \leq 1$, where $\mathcal{P}=0$ indicates that the system has not been disturbed (measured) by the measurement process, and $\mathcal{P}=1$ corresponds to performing strong or projective measurement on the system.

After the first stroke, the density operator of the system becomes as
\begin{eqnarray}
\begin{split}
\rho^{M(\mathcal{P})}&=\sum_{i=1}^{2}M_{i}\rho^{th}M_{i}^{\dagger}\\
&=\frac{1}{Z(\beta)}\bigg(\left(e^{-\beta\hbar\omega_{0}/2}+\mathcal{P}e^{\beta\hbar\omega_{0}/2}\right)|1\rangle\langle 1|+(1-\mathcal{P})e^{\beta\hbar\omega_{0}/2}|0\rangle\langle 0|\bigg),
\end{split}
\end{eqnarray} 
with average energy  
\begin{eqnarray}
\mathcal{E}^{M(\mathcal{P})}=-\frac{\hbar\omega_{0}}{2}\mathrm{tanh}(\beta\hbar\omega_{0}/2)+\mathcal{P}\hbar\omega_{0}\frac{e^{\beta\hbar\omega_{0}/2}}{Z(\beta)}.
\end{eqnarray} 
Hence, the average energy change of the working
substance due to this stroke is
\begin{eqnarray}
Q^{M(\mathcal{P})}:=\mathcal{E}^{M(\mathcal{P})}-\mathcal{E}^{th}=\mathcal{P}\hbar\omega_{0}\frac{e^{\beta\hbar\omega_{0}/2}}{Z(\beta)}.
\end{eqnarray} 
In general, $Q^{M(\mathcal{P})}\geq0$, is the amount of stochastic energy that the system obtains from the measurement process defined by operators (8). It is interesting to denote there are two important asymptotic values for the measurement strength parameter. The first case is $\mathcal{P}=1-e^{-\beta\hbar\omega_{0}}$ for which the density operator $\rho^{M(\mathcal{P})}$ has the same entropy as $\rho^{th}$, with $\mathcal{E}^{M(\mathcal{P})}=\hbar\omega_{0}\mathrm{tanh}(\beta\hbar\omega_{0}/2)/2$ and $Q^{M(\mathcal{P})}=\hbar\omega_{0}\mathrm{tanh}(\beta\hbar\omega_{0}/2)$. This means that the system gains net work (non-stochastic energy) from the measurement channel. On the other hand, for $(1-e^{-\beta\hbar\omega_{0}})/2\leq\mathcal{P}<1-e^{-\beta\hbar\omega_{0}}$, we have $S\left(\rho^{M(\mathcal{P})}\right)>S\left(\rho^{th}\right)$, i.e. the system gains stochastic energy in this way. Indeed, the second asymptotic case is related to $\mathcal{P}=\left(1-e^{-\beta\hbar\omega_{0}}\right)/2$ for which $\rho^{M(\mathcal{P})}$ becomes a completely random (mixed) state with maximal entropy, $\mathcal{E}^{M(\mathcal{P})}=0$ and $Q^{M(\mathcal{P})}=\hbar\omega_{0}\mathrm{tanh}(\beta\hbar\omega_{0}/2)/2$.

The second stroke is provided by another measurement process defined by the following operators
\begin{eqnarray}
N_{1}(\mathfrak{q})=|0\rangle\langle 0|+\sqrt{1-\mathfrak{q}}|1\rangle\langle 1|, \quad N_{2}(\mathfrak{q})=\sqrt{\mathfrak{q}}|0\rangle\langle 1|,
\end{eqnarray} 
which satisfy equation (7), and have their corresponding $POVM$ operators. In the computational basis, $N_{1}(\mathfrak{q})$ corresponds to weak measurement map on a qubit system with strength $0\leq\mathfrak{q}\leq 1$ \cite{kim,he,behzadi}. Under this stroke, the density matrix of the working substance is given by 
\begin{eqnarray}
\begin{split}
\rho^{N(\mathfrak{q})}&=\sum_{i=1}^{2}N_{i}\rho^{M}N_{i}^{\dagger}\\
&=\frac{1}{Z(\beta)}\bigg((1-\mathfrak{q})\left(e^{-\beta\hbar\omega_{0}/2}+\mathcal{P}e^{\beta\hbar\omega_{0}/2}\right)|1\rangle\langle 1|+\left((1-\mathcal{P}+\mathcal{P}\mathfrak{q})e^{\beta\hbar\omega_{0}/2}+\mathfrak{q}e^{-\beta\hbar\omega_{0}/2}\right)|0\rangle\langle 0|\bigg),
\end{split}
\end{eqnarray} 
hence, the average energy is obtained as 
\begin{eqnarray}
\mathcal{E}^{N(\mathfrak{q})}=-\frac{\hbar\omega_{0}}{2}\mathrm{tanh}(\beta\hbar\omega_{0}/2)-\mathfrak{q}\hbar\omega_{0}\frac{e^{-\beta\hbar\omega_{0}/2}+\mathcal{P}e^{\beta\hbar\omega_{0}/2}}{Z(\beta)}+\mathcal{P}\hbar\omega_{0}\frac{e^{\beta\hbar\omega_{0}/2}}{Z(\beta)}.
\end{eqnarray} 
The main demand of the scheme at this stage is that how work can be extracted from the imported energy into the engine at the first stroke. To this aim, if the measurement strength parameter in the second stroke satisfies following constraint  
\begin{eqnarray}
\mathfrak{q}=\frac{2\mathcal{P}e^{\beta\hbar\omega_{0}/2}-2\mathrm{sinh}(\beta\hbar\omega_{0}/2)}{e^{-\beta\hbar\omega_{0}/2}+\mathcal{P}e^{\beta\hbar\omega_{0}/2}},
\end{eqnarray} 
then the density matrix in (14) and the average energy in (15) will be 
\begin{eqnarray}
\rho^{N(\mathcal{P})}=\frac{1}{Z(\beta)}\bigg((1-\mathcal{P})e^{\beta\hbar\omega_{0}/2}|1\rangle\langle 1|+\left(e^{-\beta\hbar\omega_{0}/2}+\mathcal{P}e^{\beta\hbar\omega_{0}/2}\right)|0\rangle\langle 0|\bigg),
\end{eqnarray} 
and  
\begin{eqnarray}
\mathcal{E}^{N(\mathcal{P})}=\frac{\hbar\omega_{0}}{2}\mathrm{tanh}(\beta\hbar\omega_{0}/2)-\mathcal{P}\hbar\omega_{0}\frac{e^{\beta\hbar\omega_{0}/2}}{Z(\beta)},
\end{eqnarray} 
respectively. In fact, equation (16) causes the following relation 
\begin{eqnarray}
S\left(\rho^{M(\mathcal{P})}\right)=S\left(\rho^{N(\mathcal{P})}\right),
\end{eqnarray} 
i.e. under the constraint (16), the second stroke takes place in an isentropic manner so entropy of the engine does not change by the measurement channel. In addition, we observe isentropic energy reduction for the engine in the second stroke as 
\begin{eqnarray}
\Delta:=\mathcal{E}^{M(\mathcal{P})}-\mathcal{E}^{N(\mathcal{P})}=2\mathcal{P}\hbar\omega_{0}\frac{e^{\beta\hbar\omega_{0}/2}}{Z(\beta)}-\hbar\omega_{0}\mathrm{tanh}(\beta\hbar\omega_{0}/2).
\end{eqnarray} 
Therefore, the second measurement process induces unitary transformation on the working substance with average energy reduction $\Delta$, without occurring quantum friction or excitation of quantum coherences, as is clear from equation (17). In fact, excitation of quantum coherence in the working substance through the isentropic process will suppress the engine efficiency because it should be consumed an amount of work to excite quantum coherence in the working substance. This is the main result of this paper because it is shown that $\Delta$ is equal to the net work extracted from the engine. To illustrate with more explicit details, let us end the engine cycle by the third stroke provided by thermalization process. After thermalization, the exchanged thermal energy between the engine and thermal reservoir is equal to
\begin{eqnarray}
Q^{N(\mathcal{P})}:=\mathcal{E}^{th}-\mathcal{E}^{N(\mathcal{P})}=\mathcal{P}\hbar\omega_{0}\frac{e^{\beta\hbar\omega_{0}/2}}{Z(\beta)}-\hbar\omega_{0}\mathrm{tanh}(\beta\hbar\omega_{0}/2).
\end{eqnarray} 
This energy must be dissipated into the thermal reservoir, i.e. $Q^{M(\mathcal{P})}\leq0$, which leads to the constraint  $\mathcal{P}\leq1-e^{-\beta\hbar\omega_{0}}$. This is consistent with our previous consideration and confirms our design. On the other hand, the net work extracted from the engine throughout the cycle is    
\begin{eqnarray}
\begin{split}
\mathcal{W}_{\mathrm{ext}}(\mathcal{P})&=Q^{M(\mathcal{P})}+Q^{N(\mathcal{P})}\\
&=2\mathcal{P}\hbar\omega_{0}\frac{e^{\beta\hbar\omega_{0}/2}}{Z(\beta)}-\hbar\omega_{0}\mathrm{tanh}(\beta\hbar\omega_{0}/2)=\Delta.
\end{split}
\end{eqnarray}
In order to have an engine with capability of work extraction, the inequality $\mathcal{W}_{\mathrm{ext}}(\mathcal{P})\geq0$, must be satisfied. This constraint gives the lower bound  $\mathcal{P}\geq\left(1-e^{-\beta\hbar\omega_{0}}\right)/2$, for the measurement strength parameter, discussed previously. On the other hand, We find that the extracted work from the engine is equal to energy reduction of engine under constant entropy (see equation (20)), which in turns completes our main result.

Finally, to evaluate the engine efficiency, let us rewrite the measurement strength parameter in its range of variations as $\mathcal{P}=\gamma\left(1-e^{-\beta\hbar\omega_{0}}\right)$ with
\begin{eqnarray}
1/2\leq\gamma\leq1.
\end{eqnarray}
Then the efficiency becomes
\begin{eqnarray}
\eta:=\frac{\mathcal{W}_{\mathrm{ext}}(\mathcal{P})}{Q^{M(\mathcal{P})}}=2-\frac{1}{\gamma}.
\end{eqnarray}
The efficiency is zero for $\gamma=1/2$, and reaches unity for $\gamma=1$. As an illustration, the case $\gamma=1$ means that the first stroke gives energy into the engine and subsequently it is extracted completely from it by the second stroke without dissipation into thermal reservoir. In other words, during the reception of energy and extraction of it, entropy of the engine is left unchanged as recognized by equations (10) and (17). For $1/2<\gamma<1$, the first stroke gives stochastic (unorganized) energy into the engine and subsequently the organized part of that energy is extracted by the second stroke, and the remainder is dissipated into thermal reservoir. Finally for $\gamma=1/2$, the first stroke causes the density matrix of the working substance to be completely mixed such that no work can be extracted from the engine through the second stroke. Hence, all of the received energy from the first stroke is dissipated into thermal reservoir through the third stroke. In addition, let us recall the range of variation of $\gamma$ in (23) as the efficiency range of the engine, the range in which engine efficiency takes nonzero values. In the next section, when two additional adiabatic strokes are included into the engine cycle, the corresponding lower bound of efficiency range will be improved in comparison to (23). 

\section{Adiabatic extension}

In the previous section, we elucidated a qubit system with time independent Hamiltonian as working substance and obtained the equation (24) for the engine efficiency. In this section, we extend the model to include two additional strokes arise from adiabatic changing of Hamiltonian of the working substance. We consider whole of the process as 
\begin{eqnarray}
\mathrm{TP}\Longrightarrow \mathrm{API} \Longrightarrow \mathrm{QMI} \Longrightarrow \mathrm{QMII} \Longrightarrow \mathrm{ADII} \Longrightarrow \mathrm{TP}
\end{eqnarray}
where TP is the thermalization process, API and APII denote adiabatic changing the Hamiltonian (1). Also, QMI and QMII are measurement processes performed on the system through operators (8) and (13). In order to describe $\mathrm{API}$ process, we note that after isolating the engine from thermal reservoir, the transition frequency of the qubit is changed adiabatically from $\omega_{0}$ to $\omega$, then the increased level spacing of the engine is $\hbar\omega$. We note that an adiabatic process by itself takes place in an isentropic manner without occurring quantum friction. Therefore, the average energy is obtained as    
\begin{eqnarray}
\mathcal{E}^{\mathrm{API}}=-\frac{\hbar\omega}{2}\mathrm{tanh}\bigg(\frac{\beta\hbar\omega_{0}}{2}\bigg),
\end{eqnarray}
so the corresponding average work done on the engine is
\begin{eqnarray}
\mathcal{W}^{\mathrm{API}}:=\mathcal{E}^{th}-\mathcal{E}^{\mathrm{API}}=\frac{\hbar\left(\omega-\omega_{0}\right)}{2}\mathrm{tanh}\bigg(\frac{\beta\hbar\omega_{0}}{2}\bigg).
\end{eqnarray}
The engine undergoes second stroke through quantum measurement defined by operators (8). In the same way of the previous section, the related average energy constituent of the engine is 
\begin{eqnarray}
\mathcal{E}^{\mathrm{QMI}}=-\frac{\hbar\omega}{2}\mathrm{tanh}\bigg(\frac{\beta\hbar\omega_{0}}{2}\bigg)+\mathcal{P}\hbar\omega\frac{e^{\beta\hbar\omega_{0}/2}}{Z(\beta)},
\end{eqnarray}  
and the corresponding acquired stochastic energy for the engine becomes 
\begin{eqnarray}
\mathcal{Q}^{\mathrm{QMI}}:=\mathcal{E}^{\mathrm{QMI}}-\mathcal{E}^{\mathrm{API}}=\mathcal{P}\hbar\omega\frac{e^{\beta\hbar\omega_{0}/2}}{Z(\beta)}.
\end{eqnarray} 

As in the previous section, the third stroke takes place by performing general measurement according to operators (13) in an isentropic manner and without excitation of quantum coherence in the working substance. Similar to equation (18), the average energy, after this stroke, is given by  
\begin{eqnarray}
\mathcal{E}^{\mathrm{QMII}}=\frac{\hbar\omega}{2}\mathrm{tanh}\bigg(\frac{\beta\hbar\omega_{0}}{2}\bigg)-\mathcal{P}\hbar\omega\frac{e^{\beta\hbar\omega_{0}/2}}{Z(\beta)}.
\end{eqnarray} 
Hence, the isentropic energy reduction between two strokes QMI and QMII is obtained as 
\begin{eqnarray}
\Delta:=\mathcal{E}^{\mathrm{QMI}(\mathcal{P})}-\mathcal{E}^{\mathrm{QMII}(\mathcal{P})}=2\mathcal{P}\hbar\omega\frac{e^{\beta\hbar\omega_{0}/2}}{Z(\beta)}-\hbar\omega\mathrm{tanh}\bigg(\frac{\beta\hbar\omega_{0}}{2}\bigg).
\end{eqnarray} 
In the forth stroke, the Hamiltonian of the system is adiabatically returned to its original form by changing the transition frequency $\omega$ to $\omega_{0}$ and so the level spacing of the system is reduced to its original amount, i.e. $\hbar\omega_{0}$. The average energy of the system and the average extracted work from it, after this  stroke, are easily obtained as  
\begin{eqnarray}
\mathcal{E}^{\mathrm{ADII}}=\frac{\hbar\omega_{0}}{2}\mathrm{tanh}\bigg(\frac{\beta\hbar\omega_{0}}{2}\bigg)-\mathcal{P}\hbar\omega_{0}\frac{e^{\beta\hbar\omega_{0}/2}}{Z(\beta)},
\end{eqnarray} 
and 
\begin{eqnarray}
\mathcal{W}^{\mathrm{APII}}:=\mathcal{E}^{\mathrm{QMII}}-\mathcal{E}^{\mathrm{APII}}=\frac{\hbar\left(\omega-\omega_{0}\right)}{2}\mathrm{tanh}\bigg(\frac{\beta\hbar\omega_{0}}{2}\bigg)-\mathcal{P}\frac{\hbar(\omega-\omega_{0} )}{Z(\beta)}e^{\beta\hbar\omega_{0}/2},
\end{eqnarray}
respectively.
Ultimately, the engine cycle is ended by thermalization process (fifth stroke) and the exchanged stochastic energy between the working substance and the thermal reservoir is characterized as  
\begin{eqnarray}
\mathcal{Q}^{th}:=\mathcal{E}^{th}-\mathcal{E}^{\mathrm{APII}}=\mathcal{P}\hbar\omega_{0}\frac{e^{\beta\hbar\omega_{0}/2}}{Z(\beta)}-\hbar\omega_{0}\mathrm{tanh}\bigg(\frac{\beta\hbar\omega_{0}}{2}\bigg).
\end{eqnarray}
We know that for an engine this energy must be dissipated into the thermal reservoir, i.e. $\mathcal{Q}^{th}\leq0$, which gives the same constraint for the measurement strength parameter, i.e. $\mathcal{P}\leq 1-e^{-\beta\hbar\omega_{0}}$, as observed in the previous section. 
Since in the absence of any irreversible phenomena, the internal energy of the working substance is left unchanged throughout a complete cycle, then it is simply obtained that 
\begin{eqnarray}
\mathcal{Q}^{th}+\mathcal{Q}^{\mathrm{QMI}}-\mathcal{W}^{\mathrm{API}}-\Delta-\mathcal{W}^{\mathrm{APII}}=0.
\end{eqnarray}
Hence, the net work extracted from the engine throughout a complete five stroke cycle is equal to
\begin{eqnarray}
\begin{split}
\mathcal{W}_{\mathrm{ext}}(\mathcal{P})&=\mathcal{Q}^{th}+\mathcal{Q}^{\mathrm{QMI}}=\mathcal{W}^{\mathrm{API}}+\Delta+\mathcal{W}^{\mathrm{APII}}\\&=\mathcal{P}\hbar(\omega+\omega_{0})\frac{e^{\beta\hbar\omega_{0}/2}}{Z(\beta)}-\hbar\omega_{0}\mathrm{tanh}\bigg(\frac{\beta\hbar\omega_{0}}{2}\bigg).
\end{split}
\end{eqnarray}
It is expected that for this engine  $\mathcal{W}_{\mathrm{ext}}(\mathcal{P})\geq 0$, which gives the following new improved lower bound for the measurement strength parameter as 
\begin{eqnarray}
\mathcal{P}\geq \frac{\omega_{0}}{\omega_{0}+\omega}\left(1-e^{-\beta\hbar\omega_{0}}\right).
\end{eqnarray}
If there is no adiabatic process for changing the level spacing of the Hamiltonian, i.e. $\omega=\omega_{0}$, then we have the same lower bound of the previous section. For the case where
$\omega\gg\omega_{0}$,
the lower bound in (37) becomes very small. At the end, to calculate the engine efficiency in this situation, let us remember that the measurement strength parameter can be rewritten as  $\mathcal{P}=\gamma\left(1-e^{-\beta\hbar\omega_{0}}\right)$ with
\begin{eqnarray}
\omega_{0}/(\omega_{0}+\omega)\leq\gamma\leq1,
\end{eqnarray}
then the efficiency becomes
\begin{eqnarray}
\eta:=\frac{\mathcal{W}_{\mathrm{ext}}(\mathcal{P})}{\mathcal{Q}^{\mathrm{QMI}}}=1+\frac{\omega_{0}}{\omega}\bigg(\frac{\gamma-1}{\gamma}\bigg).
\end{eqnarray}
The efficiency becomes zero for $\gamma=\omega_{0}/(\omega_{0}+\omega)$, and saturates unit for $\gamma=1$. As is clear from (38), the lower bound of efficiency range becomes smaller than the case of the equation (23), so we have an improved efficiency range by introducing the adiabatic strokes.  
It is interesting to note that for $\gamma=1/2$, the efficiency is not zero and takes the following expression 
\begin{eqnarray}
\eta=1-\frac{\omega_{0}}{\omega},
\end{eqnarray}
which is the same result as obtained in \cite{talkner}. 

\section{Conclusions}

In this work, we designed a three-stroke single-qubit quantum engine with time independent working substance whose performance essentially depends on two general quantum measurements defined in (8) and (13). It was explicitly shown that the engine efficiency can be well-controlled by the measurement strength parameter such that it is enabled to reach unit. We highlighted that work extraction which takes place through the second stroke, is a frictionless isentropic process. In the next step, the engine was provided with two additional adiabatic strokes changing the transition frequency of the working substance led to improvement in lower bound of parametrized efficiency range of the engine. Particularly, it was pointed out that for $\gamma=1/2$, the introduced engine in Sec II, figured out zero efficiency while the extended engine in Sec III, rendered the same efficiency as of the engine introduced in \cite{talkner}.

\end{document}